\documentclass[aps,prb,twocolumn,showpacs]{revtex4}
\usepackage{epsfig,graphicx,longtable}
\usepackage{amsmath}
\usepackage{amsfonts}
\usepackage{amssymb}
\newcommand{\be}{\begin{equation}}
\newcommand{\ee}{\end{equation}}
\newcommand{\bi}{\bibitem}

\begin{document}

\title{Systematic investigation of a family of gradient-dependent functionals for solids}

\author{Philipp Haas}
\author{Fabien Tran}
\author{Peter Blaha}
\affiliation{Institute of Materials Chemistry, Vienna University of Technology,
Getreidemarkt 9/165-TC, A-1060 Vienna, Austria}

\author{Luana S. Pedroza}
\affiliation{Instituto de F\'\i sica, Universidade de S\~ao Paulo, Caixa Postal 
66318, S\~ao Paulo 05315-970, SP, Brazil}

\author{Antonio J. R. da Silva}
\affiliation{Instituto de F\'\i sica, Universidade de S\~ao Paulo, Caixa Postal 
66318, S\~ao Paulo 05315-970, SP, Brazil and
Laborat\'orio Nacional de Luz S\'{\i}ncrotron, Campinas, SP, Brazil}

\author{Mariana M. Odashima}
\affiliation{Instituto de F\'{\i}sica de S\~ao Carlos,
Universidade de S\~ao Paulo, S\~ao Carlos, 13560-970 S\~ao Paulo,
Brazil}

\author{Klaus Capelle}
\affiliation{Centro de Ci\^encias Naturais e Humanas,
Universidade Federal do ABC, Santo Andr\'e, 09210-170 S\~ao Paulo,
Brazil}

\begin{abstract}
Eleven density functionals are compared with regard to their performance 
for the lattice constants of solids. We consider standard functionals, such 
as the local-density approximation and the Perdew-Burke-Ernzerhof (PBE) 
generalized-gradient approximation (GGA), as well as variations of PBE GGA,
such as PBEsol and similar functionals, PBE-type functionals employing a 
tighter Lieb-Oxford bound, and combinations thereof. Several of these 
variations are proposed here for the first time. On a test set of 60 solids 
we perform a system-by-system analysis for selected functionals and a full
statistical analysis for all of them. The impact of restoring the gradient
expansion and of tightening the Lieb-Oxford bound is discussed, and 
confronted with previous results obtained from other codes, functionals 
or test sets. No functional is uniformly good for all investigated systems,
but surprisingly, and pleasingly, the simplest possible modifications to
PBE turn out to have the most beneficial effect on its performance.
The atomization energy of molecules was also considered and on a testing set
of six molecules, we found that the PBE functional is clearly the best, the
others leading to strong overbinding.
\end{abstract}

\maketitle

\section{Introduction}
\label{introduction}

Modern electronic-structure theory\cite{martin,kohanoff} relies to a very large
extent on density-functional theory (DFT).\cite{grossdreizler,parryang,kohnrmp}
The utility of DFT, in turn, depends crucially on the availability of
approximations to the exchange-correlation ($xc$) functional that are
sufficiently reliable and sufficiently simple to implement.
\cite{martin,kohanoff,grossdreizler,parryang,kohnrmp,perdewjcp}

As a consequence, a large number of approximate $xc$ functionals have been
developed. Only a few of these, however, have found widespread application,
and essentially just two of them account for the large majority of 
applications of DFT in solid-state physics: the local-density approximation
(LDA) and the Perdew-Burke-Ernzerhof (PBE) form of the generalized-gradient
approximation (GGA).\cite{pbe}

Among the main problems of these functionals is that lattice constants are 
systematically and consistently underestimated by LDA and overestimated by PBE. 
LDA lattice constants are typically about $1-5 \%$ too short, while PBE lattice 
constants are too long by almost the same margin. Many other quantities,
such as the unit-cell geometry and volume, the cohesive energy, bulk modulus,
compressibility, phonon frequencies, sound velocity, elastic constants, 
Debye temperature, the pressure-dependence of all these quantities, surface 
reconstruction energies, the possibility of structural phase transitions, 
etc., depend crucially on the lattice constant. Therefore, the difficulty 
of LDA and PBE in predicting quantitatively reliable lattice constants is 
a crucial problem standing in the way of further applications of DFT to solids. 

Until quite recently, no generally applicable solution to this problem was
in sight, and the very voluminous literature on better $xc$ functionals
(e.g., the hybrid functionals)
largely focused on finite systems (see Refs. \onlinecite{SousaJPCA07} and
\onlinecite{CramerPCCP09} for recent reviews).
For solids, however, these functionals
do not perform that well in every situation and/or lead to very expensive
calculations. For instance,
the very popular hybrid functional B3LYP\cite{b3lyp,StephensJPC94}
is rather hard to implement for solids, in particular for metals, 
and the effort does not seem to pay off, as resulting lattice constants 
overestimate experimental values by about as much as PBE.\cite{b3lypS} 
Similarly, semi-empirical\cite{pkzbmetagga} and nonempirical\cite{tpssmetagga} 
meta-GGA functionals (slightly more expensive than GGAs) produce 
little\cite{tpssmetagga,htbprb} or no \cite{pkzbmetagga} improvement for lattice
constants. Although many other functionals have been tried over the years, 
LDA and PBE remained, until very recently, the {\em de facto} standard 
DFT approach for the determination of structural properties of solids and 
nanostructures.

Recently, however, the field of functional construction for solids has gained 
new impetus through the development of AM05,\cite{am05,spinam05,MattssonJCP08} a radically
new type of density 
functional based on the subsystem approach and the Airy gas,
and the Wu-Cohen (WC) GGA,\cite{wucohen,TranPRB07} which employs a simple but
efficient modification of the 
PBE exchange enhancement factor that makes it more reliable for solid-state 
properties. An even simpler modification of PBE is PBEsol,\cite{pbesol} which 
differs from original PBE only in the values of two parameters.

These developments have rekindled the interest in the 
development of better density functionals for solids. Several such 
recently developed functionals, AM05, WC, PBEsol,
and the second-order GGA (SOGGA) of Zhao and Truhlar\cite{sogga} have been 
systematically tested and compared to LDA, PBE and TPSS meta-GGA in a 
previous publication by three of us (PH, FT, and PB).\cite{htbprb} A large test 
set (60 solids) and a very accurate all-electron implementation of the 
Kohn-Sham (KS) equations (the WIEN2k code\cite{wien}) allowed a detailed 
investigation of the performance of each of these functionals. Overall, 
no clear winner has emerged from the comparison, but the new GGA
functionals improve over LDA and PBE for many solids and give smaller
mean errors. In Ref. \onlinecite{HaasPRB09b}, we reported a detailed
analysis of the functionals, which shed light on some of the trends observed
in the lattice constants.

In an independent work, three of us (LSP, AJRdS, and KC) noted that the step that 
led from PBE to PBEsol is not unique, and allows several variations.
\cite{pbebetamu} In fact, PBE and PBEsol turned out to be just two particular members of a 
family of functionals each of which takes its parameters, $\beta$ and $\mu$,
from a different constraint. The resulting two-parameter family of functionals, 
collectively denoted PBE($\beta,\mu$), has been tested for atoms, molecules 
and solids in Ref.~\onlinecite{pbebetamu}. The calculations for solids 
performed in that work employed pseudopotentials, which is the standard 
approach for very large systems with many inequivalent sites, but introduces 
an additional source of errors and complicates an unbiased assessment of the 
performance of each functional.

In still other work, two of us (MMO and KC) initiated an investigation of 
the Lieb-Oxford (LO) bound,\cite{lo,lojcp} a fundamental property of the quantum 
mechanics of Coulomb-interacting systems according to which the exact $xc$ 
energy is bounded from below by a simple local density functional that is
proportional to the LDA for exchange. An estimate of the proportionality 
factor, $\lambda$, is a parameter in several modern density functionals, 
among them SOGGA, TPSS, WC, as well as PBE, PBEsol 
and all other members of the PBE($\beta,\mu$) family. Numerical and
analytical investigations\cite{lojcp,loijqc,lojctc,loprl} strongly suggest 
that the value adopted in standard density functionals $\lambda_{\text{LO}}=2.273$ 
is too large and should be replaced by $\lambda_{\text{EL}}=1.9555$. Of the 
functionals listed above only SOGGA makes use of this tighter bound. 
Consequences of a tighter LO bound in PBE calculations for molecular systems 
have been explored in Ref.~\onlinecite{lojctc}, but that work did not consider 
solids and did not include variations in $\beta$ and $\mu$.
We also mention the work of Peltzer y Blanc\'{a} \textit{et al}.
\cite{PeltzeryBlancaJPCM07} who concluded that reducing the value of
$\lambda$ in PBE leads to better results for the
equilibrium volume of $4d$ and $5d$ transition metals as also shown in the
present work.

In the present paper we now bring all these developments together. We propose 
and study the three-parameter family of density functionals 
PBE($\beta,\mu,\lambda$), explore all meaningful nonempirical combinations 
of these parameters that are available, implement the resulting ten functionals
in the WIEN2k all-electron code,\cite{wien} and test them on the large set 
of 60 solids from Ref.~\onlinecite{htbprb}.
In addition we implemented these functionals in the deMon code \cite{deMon} and 
tested atomization energies on a small but representative set of
six molecules.\cite{LynchJPCA03}

This paper is organized as follows. In Sec.~\ref{functionals} we describe 
the PBE($\beta,\mu,\lambda$) family of functionals, indicating the possible 
values and sources of each of its three parameters. Section~\ref{onepchanges}
is devoted to a system-by-system comparison of three members of the 
family that differ in just one constraint from original PBE. 
In Sec.~\ref{statistics} we then present results from a statistical analysis 
of the full set of ten PBE-type functionals and LDA, for all 60 solids. This 
section also contains a comparison of our results with those from several other 
published tests of similar functionals, among them various using different 
codes and different test sets. Section~\ref{classes} analyses our results 
separately for elements and compounds (metallic transition metal compounds,
semiconductors and insulators) and
Sec. \ref{molecules} reports the performance of the PBE($\beta,\mu,\lambda$)
functionals for atomization energies of small molecules.
Finally Sec. \ref{conclusions} contains our conclusions.

\section{The PBE($\beta,\mu,\lambda$) family of functionals}
\label{functionals}

The structure of PBE is explained in the original reference,\cite{pbe} 
and more details are given in the review literature.\cite{kurthperdew} 
In the interest of conciseness, we thus refrain from repeating the explicit 
expression of this widely used functional and directly focus on its 
parameters and their possible modifications.

PBE contains nonempirical parameters, whose numerical values are obtained
by requiring that the functional obeys known universal constraints. 
Two of them, $\kappa$ and $\mu$, appear in the exchange functional, $E_x$, 
and one, $\beta$, appears in the correlation functional, $E_c$. 

In the original construction of PBE,\cite{pbe} the parameter $\beta$ is chosen 
such that in the high-density limit $E_c^{\text{PBE}}$ recovers the second-order 
gradient expansion of the correlation energy of spatially weakly varying 
systems. The requirement that the combined $xc$ functional 
reproduces the LDA jellium response function (which is accurate) implies
\be
\mu = \frac{\pi^{2}}{3}\beta,
\label{LR}
\ee
which fixes $\mu$. The third parameter, $\kappa$, was determined such that 
$E_x^{\text{PBE}}$ alone obeys the Lieb-Oxford lower bound\cite{lo} on the $xc$ 
energy. This implies 
\be
\kappa = \frac{\lambda}{2^{1/3}} - 1=0.804,
\label{kappa}
\ee
where $\lambda_{\text{LO}}=2.273$ is an estimate of the Lieb-Oxford constant 
$\lambda$ obtained in Ref.~\onlinecite{lo}. 

This particular choice of constraints proved to be enormously successful, and
PBE is one of the most widely used density functionals across physics
and chemistry. Nevertheless, the choice is clearly not unique, and has recently
been reconsidered along two independent lines. To discuss these, we 
introduce the notation PBE$(\beta,\mu,\lambda)$, where the parameters can
be replaced either by their numerical values or by symbols indicating the
source of these values. Hence, original PBE becomes PBE$(G_c,J_r,\text{LO})$
indicating that $\beta$ comes from the gradient expansion of $E_c$, $\mu$
from the jellium response function, while for $\lambda$ the original
Lieb-Oxford estimate is adopted.

The first line of thought originates with the PBEsol functional, designed
specifically to improve on PBE for solids.\cite{pbesol} To construct 
PBEsol it was argued that for solids the gradient expansion of the exchange
functional is expected to be more important than that of the correlation 
functional. Consequently $\mu$, which appears in the exchange energy, is 
chosen in PBEsol such as to reproduce the second-order gradient expansion
of $E_x$. The parameter $\beta$, appearing in the correlation
energy, is determined in PBEsol by requiring that jellium surface energies
are accurately reproduced. In our notation, PBEsol becomes PBE$(J_s,G_x,\text{LO})$. 
PBEsol has been extensively tested\cite{pbesol,htbprb,csonkaetal,RopoPRB08,hydrocarbons} 
and the results have vindicated the revised choice of constraints, as PBEsol 
indeed provides significant improvement on PBE for solids (at the expense of 
worsening the results for smaller molecular systems).

Inspired by the PBEsol work, three of the present authors explored some
other possible choices of constraints for obtaining $\beta$ and $\mu$.\cite{pbebetamu}
In one of these, PBE$(G_c,G_x,\text{LO})$, $\beta$ and $\mu$ are both determined
from gradient expansions, thus guaranteeing that this expansion is
recovered, to the extent possible within the functional form of PBE, for
both exchange {\em and} correlation. In another, PBE$(J_s,J_r,\text{LO})$, $\beta$ 
and $\mu$ are both determined from jellium: $\mu$ from the jellium response 
function, as in PBE, and $\beta$ from the jellium surface energy, as in 
PBEsol. Finally, PBE$(J_r,G_x,\text{LO})$ takes $\beta$ from the jellium response
function and $\mu$ from the gradient expansion of $E_x$. The corresponding 
values of the parameters in each member of the PBE$(\beta,\mu,\text{LO})$ family are 
recorded in Table~\ref{table1}. Additional information is given in Table I
of Ref.~\onlinecite{pbebetamu}.

\begin{figure*}
\includegraphics[height=185mm,width=80mm,angle=-90]{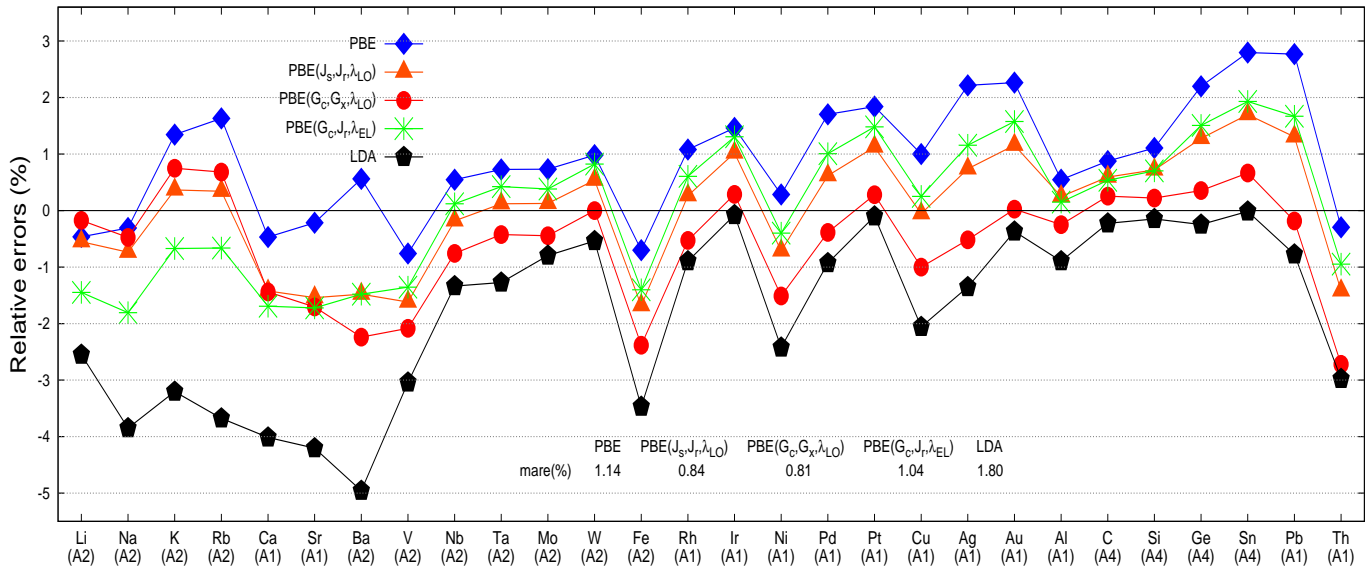}
\caption{\label{fig1} (Color online)
Relative error in the lattice constants of 28 elemental solids, obtained from
LDA, original $\text{PBE}=\text{PBE}(G_c,J_r,\text{LO})$ and three versions
of PBE that differ from it in just one constraint each, as described in the main text.
Inset: Mean absolute relative error (mare) of the five functionals in the
figure on this set of elemental solids.
The Strukturbericht symbols (in parenthesis) are used for the
structure: A1 = fcc, A2 = bcc, and A4 = diamond.}
\end{figure*}

{\em A priori} one might expect that functionals such as PBE$(G_c,G_x,\text{LO})$
and PBE$(J_s,J_r,\text{LO})$ that take $\beta$ and $\mu$ from the same type of
source, have the potential to benefit from error cancellation between the
exchange and the correlation functional to a larger extent than functionals
such as PBE and PBEsol that take them from different types of source.
Also, one might anticipate that PBE$(J_s,J_r,\text{LO})$ should be rather good for
simple metals, as its takes both of its parameters from jellium, the 
paradigmatic model of such metals. These expectations were put to the test 
in Ref.~\onlinecite{pbebetamu}, for atoms, molecules and solids. For each 
class of systems, a different ranking of functionals was found. Here we
only record that for solids, where the calculations were done with the
Siesta code,\cite{siesta} original PBE performed worst of all. As
expected, PBEsol provided significant improvement on PBE, but in spite
of its name and the rationale behind its construction, it did not
consistently provide the best performance for solids. Rather, best 
lattice constants were obtained from PBE$(G_c,G_x,\text{LO})$. It was not clear,
however, to which extent this conclusion was affected by the pseudopotential
approximation and the special basis functions employed in the Siesta code.

In a second, independent, line of thought, the role of the Lieb-Oxford
bound in functional construction has recently been reconsidered. Initial
numerical and analytical evidence\cite{lojcp,loijqc,lojctc} suggested that 
the Lieb-Oxford estimate $\lambda_{\text{LO}}=2.273$ could be tightened to a value 
close to $\lambda \approx 2$. Later, general arguments were given\cite{loprl} 
that for three-dimensional systems this value should actually be 
$\lambda_{\text{EL}}=1.9555$, where the subscript EL indicates that this is the 
exact value in the low-density limit of the electron liquid. This reduced 
value of $\lambda$ implies a corresponding reduction of $\kappa$ to
$0.552$. In our present notation, the resulting functional is denoted 
PBE$(G_c,J_r,\text{EL})$, and differs from original PBE only in the value of
$\lambda$ (or, equivalently, $\kappa$). This functional has been
tested for a variety of molecular 
systems\cite{lojctc} and it was found that PBE is rather 
insensitive to changes in $\lambda$ for covalently and ionically bound small 
molecules, a reduced, and thus, in principle, better, value of $\lambda$ 
producing slightly worsened energies and slightly improved bond lengths.

In the present work we now tie up various open ends from these
previous investigations, by implementing all ten functionals that can be
obtained from the above-described combinations of $\beta$, $\mu$ and 
$\lambda$, i.e., the complete family PBE$(\beta,\mu,\lambda)$, in
the all-electron code WIEN2k,\cite{wien} and testing them systematically
for a large set of 60 solids, comprising metals, semiconductors and 
insulators.\cite{htbprb} The PBE$(\beta,\mu,\lambda)$ functionals were also
tested on a set of six molecules for the atomization energy.
This test set (called AE6) was
proposed by Lynch and Truhlar \cite{LynchJPCA03} as a representative
set of a much larger set of molecules. The molecules in the AE6 set are
SiH$_{4}$, SiO, S$_{2}$, C$_{3}$H$_{4}$, C$_{2}$H$_{2}$O$_{2}$,
and C$_{4}$H$_{8}$.

The calculations on solids were performed with the WIEN2K code
\cite{wien} which solves the 
KS equations using the full-potential (linearized) augmented plane-wave 
and local orbitals [FP-(L)APW+lo] method.\cite{Singh}
Because the FP-(L)APW+lo method 
is one of the most accurate methods to solve the KS equations it represents 
a good choice for testing $xc$ functionals. The error in a 
calculated ground-state property is solely due to the approximate 
functional if good convergence parameters have been used. All calculations 
have been converged with respect to the number of $\mathbf{k}$-points and the
size of the basis set. Spin-orbit coupling for solids containing Ba, Ce, 
Hf, Ta, W, Ir, Pt, Au, Pb, and Th atoms has been taken into account.
The experimental lattice constants are taken from Ref.~\onlinecite{htbprb} 
and are corrected for zero-point anharmonic expansion.
The calculations on molecules were done with the deMon code \cite{deMon}
which uses Gaussian basis sets. The very large
uncontracted basis sets developed by Partridge \cite{PartridgeJCP87,PartridgeJCP89}
were used.

The statistical quantities that will be used for the analysis are displayed below,
where $p_{i}^{\text{calc}}$ and $p_{i}^{\text{exp}}$ are the calculated and
experimental values of the considered property (either the lattice constant
or the atomization energy) of the $i$th solid or molecule of the testing set:

mean error (in \AA~or in kcal/mol), 
\be
\text{me} = \frac{1}{n}\sum\limits_{i=1}^n
\left(p_{i}^{\text{calc}}-p_{i}^{\text{exp}}\right),
\label{me}
\ee
the mean absolute error (in \AA~or in kcal/mol), 
\be
\text{mae} = \frac{1}{n}\sum\limits_{i=1}^n
\left\vert p_{i}^{\text{calc}}-p_{i}^{\text{exp}}\right\vert,
\label{mae}
\ee
the mean relative error (in \%), 
\be
\text{mre} = \frac{1}{n}\sum\limits_{i=1}^n
100\frac{p_{i}^{\text{calc}}-p_{i}^{\text{exp}}}{p_{i}^{\text{exp}}},
\label{mre}
\ee
and the mean absolute relative error (in \%),
\be
\text{mare} = \frac{1}{n}\sum\limits_{i=1}^n
100\left\vert\frac{p_{i}^{\text{calc}}-p_{i}^{\text{exp}}}{p_{i}^{\text{exp}}}\right\vert.
\label{mare}
\ee
The spread (in \%), defined as 
\begin{eqnarray}
\text{spread} & = &
\max\left(100\frac{p_{i}^{\text{calc}}-p_{i}^{\text{exp}}}{p_{i}^{\text{exp}}}\right) \nonumber \\
& & -\min\left(100\frac{p_{i}^{\text{calc}}-p_{i}^{\text{exp}}}{p_{i}^{\text{exp}}}\right)
\label{spread}
\end{eqnarray}
will also be discussed.
The smaller the spread, the more predictable a
functional behaves. A large spread, by contrast, indicates a more erratic behaviour.
In situations where a single bad value can be problematic, it may
be wiser to choose a functional with a small spread than one with a low
mean error, as the latter may still be way off in isolated cases.

\section{Analysis of single-constraint changes with respect to PBE}
\label{onepchanges}

The calculated lattice constants for all functionals considered in this work
are given in Table SI of the supplementary EPAPS material.\cite{EPAPS}
Graphical representations of these results are also given in Figs. S1, S2, 
and S3.

In a first step, we focus our analysis on a subset of functionals that
differ from original PBE in the choice of just one constraint. Specifically,
these functionals are PBE$(J_s,J_r,\text{LO})$, differing in the
constraint for determining $\beta$, PBE$(G_c,G_x,\text{LO})$,
differing in the constraint chosen for $\mu$, and 
PBE$(G_c,J_r,\text{EL})$, differing in the choice of $\lambda$. Initial focus 
on just these functionals is useful because it allows to separately assess 
the influence of each change relative to PBE. 

\begin{figure*}
\includegraphics[height=185mm,width=80mm,angle=-90]{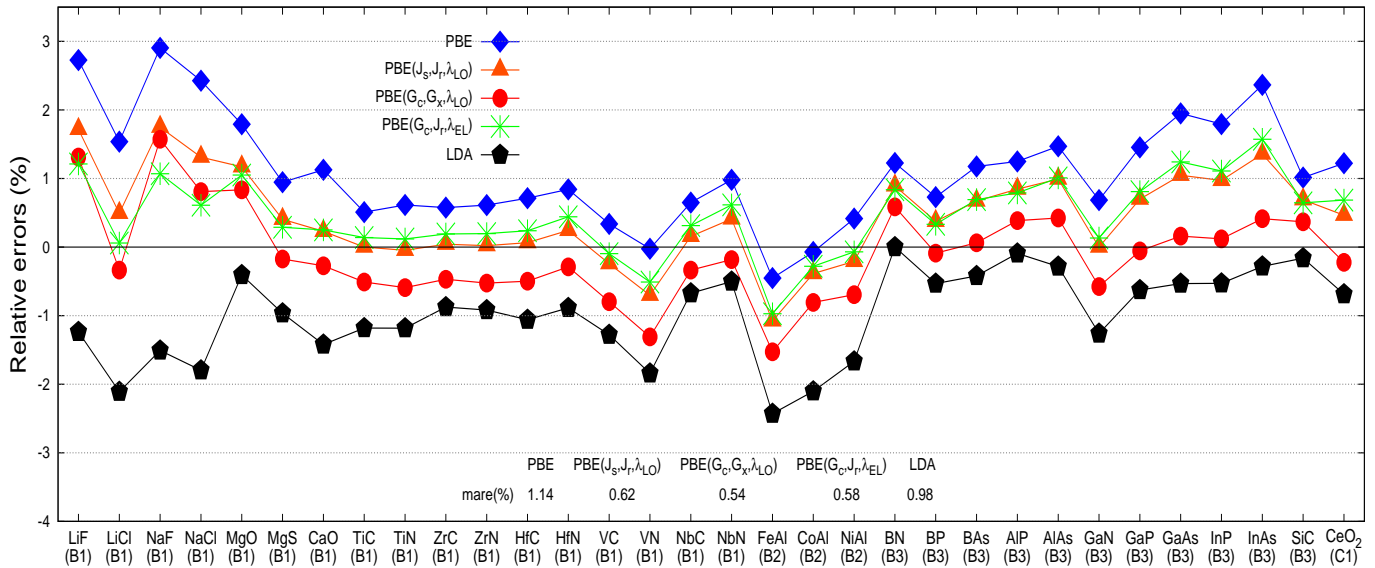}
\caption{\label{fig2} (Color online) Relative error in the lattice constants
of 32 compound solids, obtained from LDA, original
$\text{PBE}=\text{PBE}(G_c,J_r,\text{LO})$
and three versions of PBE that differ from it in just one constraint each,
as described in the main text.
Inset: Mean absolute relative error (mare) of the five functionals in the
figure on this set of compound solids.
The Strukturbericht symbols (in parenthesis) are used for the
structure: B1 = rock-salt,
B2 = cesium-chloride, B3 = zinc-blende, and C1 = fluorite.}
\end{figure*}

Figures \ref{fig1} and \ref{fig2} show the signed relative errors for all 
60 solids, for the three functionals just described as well as for original 
PBE and LDA. While original PBE has the same (modest) performance for 
elemental solids as for compounds, all other functionals work better for
compounds. The difference is particularly pronounced for LDA and 
PBE$(G_c,J_r,\text{EL})$, which perform significantly better for compounds than
for elemental solids. In both classes, however, PBE$(G_c,G_x,\text{LO})$ achieves
the lowest mare among this subset of functionals.

Figures \ref{fig1} and \ref{fig2} also show very clearly the known trend of
LDA to underestimate the lattice
constants (negative relative errors) and of PBE to overestimate the lattice
constants (positive relative errors). The most interesting fact, which is not
at all obvious from the way the various functionals were constructed, is that
{\em all changes of parameters relative to PBE produce significantly better 
lattice constants.} (This remains true even if all the other possible 
combinations are included.) 
Since the nature and source of each modified parameter are completely 
different in each of the three cases, this seems to indicate that the original 
PBE choice was rather unfortunate for lattice constants, as reasonable changes
to any of its parameters end up improving the results. 

Comparing the absolute size of the change resulting from each modified
parameter, we immediately conclude from Figs. \ref{fig1} and \ref{fig2} that PBE is most 
sensitive to changes in $\mu$ and least sensitive to changes in 
$\lambda$.\cite{footnote1}
The relative impact of $\beta$ and $\mu$ is quite reasonable, because 
$\beta$ appears in the correlation energy and $\mu$ in the exchange energy. 
Since the exchange energy in ordinary solids is larger than the correlation 
energy the result should indeed depend more sensitively on changes of that 
quantity. 

Regarding changes only in $\lambda$, we see that a change from the 
Lieb-Oxford value to the electron-gas value has, for most solids, the smallest 
effect on PBE of the three tested parameter changes. This is consistent with 
what was previously observed for molecules.\cite{lojctc} The exception 
is for the alkali metals, where this change has the biggest effect. The
reasons for this behaviour becomes clear from the analysis given by
Haas \textit{et al}.
(Ref. \onlinecite{HaasPRB09b}). In this paper an ``important region'' was defined which is
to a large extent responsible for the changes in lattice parameters of different
functionals. For
closed packed solids (like elements in the fcc or bcc structure) this region is the separation
between the outermost core (``semi-core'') and the valence states and for the
alkali metals the reduced density gradient
$s=\left\vert\nabla\rho\right\vert/\left(2\left(3\pi^{2}\right)^{1/3}\rho^{4/3}\right)$
in this ``important region'' is
much larger (even above $s=2$ for Li) than for other elements (e.g., in bcc V,
$s_{\text{max}}=0.9$). Obviously, a change in $\lambda$ modifies the enhancement
factor $F_{xc}(r_{s},s)$
much more for large $s$, while a change in $\mu$ influences $F_{xc}(r_{s},s)$
predominantly in the low $s$ region.

\section{Full statistics for all ten PBE$(\beta,\mu,\lambda)$ functionals}
\label{statistics}

In this section we present a statistical analysis of all ten functionals of
the PBE$(\beta,\mu,\lambda)$ family, as well as of LDA. We do not include
numerical data on other GGA-type functionals, such as SOGGA\cite{sogga} and
WC\cite{wucohen} and neither on alternative functionals such as
AM05\cite{am05} and TPSS meta-GGA,\cite{tpssmetagga} as these were already 
investigated on the same test set in Ref.~\onlinecite{htbprb}, and on a smaller 
set (using different codes and basis sets)
in Refs.~\onlinecite{MattssonJCP08,RopoPRB08,csonkaetal}. 
However, in our discussion of global trends we compare with results and
conclusions from those references. We do not consider earlier variations of 
PBE, such as revPBE (proposed in the Comment of Zhang and Yang\cite{pbe})
and RPBE \cite{RPBE} which were not designed 
for solids and tend to worsen PBE for extended systems
(see, e.g., Ref. \onlinecite{MadsenPRB07} for RPBE).

The choice of the best performing functional for lattice constants depends on
the measure of error selected to judge the performance of the functionals.
In terms of the mean error, PBE$(J_r,G_x,\text{LO})$ achieves a spectacularly low
error of $-0.002$ \AA, followed by PBEsol and PBE$(J_s,J_r,\text{EL})$. These
same three functionals also achieve the best mean relative errors. However,
such low errors can in part be due to the result of error cancellation in taking the averages.
The mean absolute error, which is not influenced by error cancellation,
favours PBEsol, closely followed by PBE$(G_c,G_x,\text{LO})$ and PBE$(J_r,G_x,\text{LO})$.
The same three functionals also achieve the lowest mean absolute relative
error. 

In view of the very small differences in the mares of some of the functionals, one 
must ask how significant these differences are, considering both, the numerical
accuracy of the theoretical results and the 
accuracy of the experimental data, where not for all cases high quality 
low-temperature lattice parameters and good zero-point energy corrections are 
available. In fact, changes in mare of 0.05\% are about the limit of theoretical 
accuracy, i.e., we expect to have an absolute precision of about 
$0.005$ \AA. Therefore, we consider PBEsol, PBE$(J_r,G_x,\text{LO})$ and 
PBE$(G_c,G_x,\text{LO})$ to perform equally well in terms of the mare.

Three of the four functionals with good mre or mare take the original
$\lambda_{\text{LO}}$ and $\mu$ from the gradient expansion for 
exchange ($G_x$), but differ in the parameter of the correlation functional, 
$\beta$. As the numbers show, the value of $\beta$ turns out to be almost 
irrelevant, while the value of $\mu$ appears to be responsible for the
improved behaviour. The gradient expansion for exchange is thus seen 
to be the key ingredient in functionals that deliver good lattice constants.
When $\mu$ is taken a bit larger (e.g., $J_r$), both $\beta$ and $\lambda$ must be set
to small values ($\beta=J_s$ and $\lambda=\lambda_{\text{EL}}$) to obtain a similar performance
[PBE$(J_s,J_r,\text{EL})$].
This observation, in retrospect, vindicates the PBEsol approach\cite{pbesol} 
and attests to the solidity of the basic insight presented there regarding 
the relevance of the gradient expansion of the exchange energy for solids.

\begin{table*}
\caption{\label{table1} Parameters of the eleven functionals under investigation
in this work, and a statistical summary [Eqs. (\ref{me})-(\ref{spread})] of their performance for lattice constants
of 60 solids. Lattice constants differing by less than $\approx 0.005$ \AA\,
and mares differing by less than $\approx 0.05\%$ should be considered
equivalent.
The lattice constants are given in Table SI of the supplementary EPAPS material.\cite{EPAPS}}
\begin{tabular}{lcccccccc}
\hline
\hline
Functional                               & $\beta$ & $\mu$ & $\lambda$\footnotemark[5] & me (\AA) & mae (\AA) & mre (\%) & mare (\%) & spread (\%) \\
\hline
LDA                                      & -     & -       & -      & $-0.060$ & 0.060 & $-1.37$ & 1.37 & 4.95 \\
PBE$(G_c,J_r,\text{LO})$\footnotemark[1] & 0.067 & 0.21951 & 2.273  &   0.049  & 0.053 &   1.02  & 1.14 & 3.68 \\
PBE$(J_s,G_x,\text{LO})$\footnotemark[2] & 0.046 & 0.12346 & 2.273  & $-0.007$ & 0.028 & $-0.21$ & 0.64 & 3.71 \\
PBE$(J_s,J_r,\text{LO})$\footnotemark[3] & 0.046 & 0.15133 & 2.273  &   0.014  & 0.033 &   0.25  & 0.72 & 3.41 \\
PBE$(G_c,G_x,\text{LO})$\footnotemark[3] & 0.067 & 0.12346 & 2.273  & $-0.012$ & 0.029 & $-0.31$ & 0.67 & 4.30 \\
PBE$(J_r,G_x,\text{LO})$\footnotemark[3] & 0.038 & 0.12346 & 2.273  & $-0.002$ & 0.030 & $-0.09$ & 0.67 & 3.56 \\
PBE$(G_c,J_r,\text{EL})$\footnotemark[4] & 0.067 & 0.21951 & 1.9555 &   0.014  & 0.037 &   0.28  & 0.80 & 3.73 \\
PBE$(J_s,G_x,\text{EL})$                 & 0.046 & 0.12346 & 1.9555 & $-0.023$ & 0.033 & $-0.55$ & 0.76 & 4.00 \\
PBE$(J_s,J_r,\text{EL})$                 & 0.046 & 0.15133 & 1.9555 & $-0.008$ & 0.032 & $-0.20$ & 0.71 & 3.87 \\
PBE$(G_c,G_x,\text{EL})$                 & 0.067 & 0.12346 & 1.9555 & $-0.028$ & 0.034 & $-0.67$ & 0.78 & 3.75 \\
PBE$(J_r,G_x,\text{EL})$                 & 0.038 & 0.12346 & 1.9555 & $-0.018$ & 0.034 & $-0.44$ & 0.75 & 4.08 \\
\hline
\hline
\footnotetext[1]{This is PBE.\cite{pbe}}
\footnotetext[2]{This is PBEsol.\cite{pbesol}}
\footnotetext[3]{These are the three functionals proposed in Ref.~\onlinecite{pbebetamu}.}
\footnotetext[4]{This is the functional of Ref.~\onlinecite{lojctc}.}
\footnotetext[5]{$\lambda=2.273$ and 1.9555 correspond to $\kappa=0.804$ and 0.552, respectively
[see Eq. (\ref{kappa})].}
\end{tabular}
\end{table*}

In terms of the spread [Eq. (\ref{spread})], the best performer is PBE$(J_s,J_r,\text{LO})$, followed by 
PBE$(J_r,G_x,\text{LO})$. 
Unfortunately, the best performers with regard to mean errors and with regard 
to the spread are not always the same. In particular PBE$(G_c,G_x,\text{LO})$ has a
rather large spread, but PBE$(J_r,G_x,\text{LO})$ appears to be a 
reasonable compromise, doing well according to all three 
criteria.

Next, we compare our present conclusions to several different sets
of earlier calculations, testing specific members of the full family,
employing other test sets, or other implementations and basis functions.

In Ref.~\onlinecite{pbebetamu}, three of us tested the five functionals
PBE$(\beta,\mu,\text{LO})$ on a set of 13 solids. These functionals were implemented
in the Siesta code.\cite{siesta} This code, by design, aims at the electronic
structure of very large systems, where all-electron calculations, even with
simple functionals, would be prohibitively expensive. To this end, it makes 
use of specially designed localized numerical basis functions, and 
pseudopotentials. As a consequence, it does not attain the same high accuracy
as all-electron codes, such as WIEN2k,\cite{wien} 
and the absolute size of the errors is larger for Siesta than 
for WIEN2k. 

Nevertheless, the resulting error statistics is rather similar (although not
identical). In particular, both the Siesta and the WIEN2k calculations identify
original PBE and PBE$(J_s,J_r,\text{LO})$ as the worst performers for lattice
constants of all PBE($\beta,\mu,\text{LO}$) functionals, and PBEsol,
PBE$(J_r,G_x,\text{LO})$, and PBE$(G_c,G_x,\text{LO})$ tied as the best.
Among these best performing functionals, Siesta and WIEN2k produce a
different ranking, with Siesta preferring PBE$(G_c,G_x,\text{LO})$, which according 
to WIEN2k is beaten by a small margin by PBEsol and PBE$(J_r,G_x,\text{LO})$. 

Interestingly, all-electron calculations for solids performed with the 
Gaussian code\cite{gaussian} also indicate that PBE$(G_c,G_x,\text{LO})$ produces 
better lattice constants than PBEsol
(see Table SIV in the supplementary EPAPS material of 
Ref.~\onlinecite{pbesol}.) which is in line 
with the Siesta results. However since neither pseudopotentials (Siesta) nor 
Gaussian basis functions (Gaussian) are as accurate for solids as all-electron 
FP-(L)APW+lo calculations, and since all differences are rather small, we still regard 
these functionals as essentially tied.

In Ref.~\onlinecite{lojctc}, two of us with Samuel B. Trickey tested the 
functional PBE$(G_c,J_r,\text{EL})$, which differs from original PBE 
only in the reduction of $\lambda$, corresponding to a tighter (and thus 
presumably better) Lieb-Oxford bound. The calculations were done for a set
of small molecules.
The reduction of $\lambda$ was found to slightly improve interatomic 
distances. In parallel, from Siesta pseudopotential calculations \cite{Pedroza} for the 13 solids 
of Ref.~\onlinecite{pbebetamu} we found that the same 
improvement occurs also for the other members of the PBE$(\beta,\mu,\lambda)$ 
family, all of which produce better lattice constants when a tighter 
Lieb-Oxford bound is enforced. By contrast, in the WIEN2k calculations only 
the badly performing functionals [original PBE and PBE$(J_s,J_r,\text{LO})$] benefit 
from a reduced value of $\lambda$, while the mare of the other three 
functionals grows if $\lambda$ is reduced. Consistently with what was 
speculated in Ref.~\onlinecite{lojctc}, this indicates that the algebraic 
form of PBE is too restricted to systematically benefit from a tighter bound.

In Ref.~\onlinecite{htbprb}, three of us employed the same set of
60 solids to assess the performance of the AM05, WC, PBEsol, SOGGA, and the
meta-GGA TPSS
functionals compared to the older LDA and PBE.
From this comparison PBEsol emerged as the 
functional with the lowest mare over all 60 solids, tied with the
WC, closely followed by SOGGA and AM05, and more distantly by 
TPSS, PBE and LDA, in this order. WC, SOGGA, and
AM05 turn out to have mares in the same range as the members of the 
PBE$(\beta,\mu,\lambda)$ family (including PBEsol) although they differ 
from the original PBE by more than just the values of parameters.

The hybrid functional B3LYP,\cite{b3lyp,StephensJPC94} which is very popular in quantum
chemistry, was shown\cite{b3lypS} to overestimate lattice constants by at 
least as much as PBE, and is thus not competitive with any of the
functionals under study here.

Interestingly, SOGGA turns out to be a very good functional making
use of the tighter Lieb-Oxford bound, using $\lambda_{\text{EL}}$ instead of
$\lambda_{\text{LO}}$. SOGGA achieves a lower mare (0.68\%) than any of the
five functionals PBE$(\beta,\mu,\text{EL})$, whose mare ranges from 
0.71 to 0.80\%, but unfortunately its mre is twice as large as that of 
PBE$(J_s,J_r,\text{EL})$. This indicates, one more time, that the functional form 
of PBE must be changed to fully benefit from a tighter Lieb-Oxford bound,
and hints that the form of SOGGA may be a suitable starting point for
this purpose.

\section{\label{classes}Analysis for classes of systems}

The above discussion was based on the statistical data of 
the full set of 60
solids. However, it can also be interesting to analyze the results
for a particular class of solids, therefore, below we
discuss the performance of all 11 functionals separately for certain
classes of solids, for which
Table SI and Fig. S1 of the supplementary EPAPS material\cite{EPAPS} are
useful.

\subsection{\label{ElSolids} Elemental solids}

Let us start out the discussion with the alkali metals.
For a given $\beta$ and $\mu$, the reduction of $\lambda$ from LO to
EL leads to significantly smaller lattice constants $a_{0}$. As mentioned before, this effect
is particularly strong for the alkali metals 
and thus all PBE$(\beta,\mu,\lambda)$
functionals with a tighter Lieb-Oxford bound underestimate the lattice
constant. Actually, from Fig. \ref{fig1} and S1 we can see that for the alkali metals
(and also the alkali-earth metals and the compounds with these elements)
the difference between the PBE ($\lambda=\lambda_{\text{LO}}$) and
LDA ($\lambda=1$) relative errors is large, an effect which is (at least partially)
due to the large values of $s$ (which make $\lambda$ important) in the region of separation between semi-core and
valence electrons.\cite{HaasPRB09b}
By comparing the results with fixed $\mu=G_x$ and a variation of $\beta$ [PBE$(J_s,G_x,\text{\text{LO/EL}})$ with
PBE$(G_c,G_x,\text{\text{LO/EL}})$ and PBE$(J_r,G_x,\text{LO/EL})$] we note
that an increase of $\beta$ from $J_s$ to $G_c$
increases the lattice constant more than a reduction of $J_r$ decreases it. $J_r$ worsens
the underestimation of $a_{0}$ from PBE$(J_s,G_x,\text{LO/EL})$ while
$G_c$ reduces it. This trend, namely that $a_{0}$(PBE$(G_c,G_x,\text{LO/EL})$) $>$
$a_{0}$(PBE$(J_s,G_x,\text{LO/EL})$) $>$ $a_{0}$(PBE$(J_r,G_x,\text{LO/EL})$) can also be observed
for most group IIA elements, but in all other cases the trend is inverted.
On the other hand, when $\mu$ is taken as $J_r$ [and therefore depends on
$\beta$, see Eq. (\ref{LR})], an increase of $\beta=J_s$ to $\beta=G_c$
[comparing PBE$(J_s,J_r,\text{LO/EL})$ with 
PBE$(G_c,J_r,\text{LO/EL})$] leads to increased lattice parameters for all
classes of compounds, not just the group IA and IIA elements. This effect is
most pronounced in K and Rb and reduces the large overestimation
of PBE$(G_c,J_r,\text{LO})$ for these two compounds.
Usually, a reduction of $\mu = J_r$ to $\mu = G_x$ (at any $\beta$ and $\lambda$)
[PBE$(G_c,J_r,\text{LO/EL})$ and PBE$(G_c,G_x,\text{LO/EL})$] leads to
much smaller lattice constants, but for group IA elements
this general trend is not true for Li (and in very few cases for Na and K).
In general, for group IA elements PBE$(J_s,G_x,\text{LO})$ is the most accurate
PBE$(\beta,\mu,\lambda)$ functional followed by PBE$(G_c,G_x,\text{LO})$.

For the elements of group IIA the original PBE [PBE$(G_c,J_r,\text{LO})$] gives
already quite satisfactory results and thus all modifications of the 
PBE$(\beta,\mu,\lambda)$ functionals lead to strong underestimations of the
lattice constants. Tightening the Lieb-Oxford
bound increases the absolute relative error by about $1.5 \%$. Reduction of
$\mu$ has an even larger negative effect. Changing $\beta$ (at fixed $\mu = G_x$) 
has a fairly small effect (in particular for Ba) and a non-uniform trend.

The lattice parameters of group IVA elements are very well described by
standard LDA (only for Pb there is a significant underestimation), while the
original PBE functional yields $a_0$ almost $3 \%$ too large for Sn and Pb. In
addition, PBE shows a strong tendency for larger overestimation of $a_0$ for heavier
elements.
Since LDA is so good, only the ``weakest'' GGAs, i.e., where the exchange
parameter $\mu$ is small ($G_{x}$), can compete with LDA. When in addition also
the reduced $\lambda_{\text{EL}}$ is used and/or $\beta = G_c$ is kept large, functionals
like PBE$(G_c,G_x,\text{EL})$, PBE$(J_s,G_x,\text{EL})$ or
PBE$(G_c,G_x,\text{LO})$ (in that order) perform very well.

The trend for the $3d$ transition metals (TM) is very similar to that for the group
IIA elements. Since already original PBE is rather accurate
for the $3d$ TM (except for Cu), no overall improvement can be
expected when one (or several) of the parameters is reduced.
A tighter Lieb-Oxford bound [PBE$(G_c,J_r,\text{EL})$]
improves the situation for Cu, but worsens all
other cases. A reduction of $\mu$ has an even larger negative effect, but
PBE$(J_s,J_r,\text{LO})$ is the best modified $\lambda = \lambda_{\text{LO}}$ functional. None of
the modifications seems to be able to break the trend that lattice parameters
of early $3d$ TM are even more underestimated than of late ones.
 
The lattice constants of the $4d$ transition metals are overestimated by
original PBE (the overestimation increases for later TM) and slightly underestimated
by PBEsol (getting more accurate with increasing nuclear charge). 
Using a tighter Lieb-Oxford
bound ($\lambda_{\text{EL}}$) therefore improves PBE (worsens PBEsol), but this
modification alone is not enough. An additional reduction of $\beta$
from $G_c$ to $J_s$ [PBE$(J_s,J_r,\text{EL})$, which also reduces the effective $\mu$]
leads to pretty
accurate results, while the $\beta$ reduction alone [PBE$(J_s,J_r,\text{LO})$] is not sufficient. 
Similar good results can also be reached when both, $\mu$ and $\beta$ are
reduced to $G_x$ and $J_r$, respectively [PBE$(J_r,G_x,\text{LO})$]. Most
interestingly, these two modifications can also significantly reduce the trend towards larger
lattice parameters for later TM and are thus an improvement for \textit{all} $4d$ elements. 
Any further combination with reduced $\beta, \mu$, and $\lambda$ underestimates
the lattice parameters. 

For the $5d$ TM, original PBE overestimates $a_{0}$ and for the latest $5d$
element (Au), the error reaches more than $2 \%$. The best functional
for the $5d$ elements is PBE$(G_c,G_x,\text{LO})$, where only the exchange factor
$\mu$ is strongly reduced, but $\beta$ (and $\lambda$) are kept at the large
values. As for the $4d$ series, the trend towards larger lattice parameters for
late TM elements is more or less completely broken. 
Similar, but slightly overestimated $a_0$ can be obtained when both, $\beta$
and $\lambda$ are also reduced [PBE$(J_s,G_x,\text{EL})$]. Reduction of
$\lambda$ alone or intermediate values for $\mu$ still overestimate the lattice
parameters.

For the heaviest element of our testing set, the $5f$ element Th, the original
PBE gives the best result (still underestimating $a_0$ slightly), while,
e.g., PBEsol leads to a more than $2 \%$ too small lattice parameter.

\begin{table*}
\caption{\label{table2} Parameters of the eleven functionals under investigation
in this work, and a statistical summary [Eqs. (\ref{me})-(\ref{mare})] of their performance for the atomization energy
of the set AE6 of six molecules.\cite{LynchJPCA03}
The atomization energies are given in Table SII of the supplementary EPAPS material.\cite{EPAPS}}
\begin{tabular}{lccccccc}
\hline
\hline
Functional                               & $\beta$ & $\mu$ & $\lambda$\footnotemark[5] & me (kcal/mol) & mae (kcal/mol) & mre (\%) & mare (\%) \\
\hline
LDA                                      & -     & -       &    -  & 76.3 & 76.3 & 16.9 & 16.9 \\ 
PBE$(G_c,J_r,\text{LO})$\footnotemark[1] & 0.067 & 0.21951 & 2.273 & 12.0 & 15.1 &  3.4 &  4.4 \\ 
PBE$(J_s,G_x,\text{LO})$\footnotemark[2] & 0.046 & 0.12346 & 2.273 & 35.1 & 35.1 &  8.3 &  8.3 \\ 
PBE$(J_s,J_r,\text{LO})$\footnotemark[3] & 0.046 & 0.15133 & 2.273 & 28.5 & 28.7 &  6.9 &  6.9 \\ 
PBE$(G_c,G_x,\text{LO})$\footnotemark[3] & 0.067 & 0.12346 & 2.273 & 31.0 & 32.7 &  7.6 &  8.2 \\ 
PBE$(J_r,G_x,\text{LO})$\footnotemark[3] & 0.038 & 0.12346 & 2.273 & 36.4 & 36.4 &  8.5 &  8.5 \\ 
PBE$(G_c,J_r,\text{EL})$\footnotemark[4] & 0.067 & 0.21951 & 1.9555 & 26.6 & 28.5 &  6.5 &  7.1 \\ 
PBE$(J_s,G_x,\text{EL})$                 & 0.046 & 0.12346 & 1.9555 & 43.5 & 43.5 & 10.1 & 10.1 \\ 
PBE$(J_s,J_r,\text{EL})$                 & 0.046 & 0.15133 & 1.9555 & 38.9 & 38.9 &  9.1 &  9.1 \\ 
PBE$(G_c,G_x,\text{EL})$                 & 0.067 & 0.12346 & 1.9555 & 39.4 & 40.4 &  9.4 &  9.7 \\ 
PBE$(J_r,G_x,\text{EL})$                 & 0.038 & 0.12346 & 1.9555 & 44.8 & 44.8 & 10.3 & 10.3 \\ 
\hline
\hline
\footnotetext[1]{This is PBE.\cite{pbe}}
\footnotetext[2]{This is PBEsol.\cite{pbesol}}
\footnotetext[3]{These are the three functionals proposed in Ref.~\onlinecite{pbebetamu}.}
\footnotetext[4]{This is the functional of Ref.~\onlinecite{lojctc}.}
\footnotetext[5]{$\lambda=2.273$ and 1.9555 correspond to $\kappa=0.804$ and 0.552, respectively
[see Eq. (\ref{kappa})].}
\end{tabular}
\end{table*}

\subsection{\label{compounds} Compounds}

Most prior discussed trends (Sec. \ref{ElSolids}) can to some extent also be
observed for compounds. Sometimes the combined effect of two elements may lead to
some kind of cancellation, or in other cases, one element may dominate the
effect. We will discuss below the effects starting with the very ionic group I-VII
and II-VI compounds, then covering the more covalently bound group III-V and
TM-compounds.

For the ionic compounds it is obvious that the anion (the tails of the valence
$p$ electron density) plays the major role in
determining the lattice parameter. This was shown in Ref. \onlinecite{HaasPRB09b}, but
is also obvious by comparing the lattice parameters for e.g. metallic Li
and LiF. Using PBE there is a small underestimation of $a_0$ for Li, but the lattice
parameter of LiF is too large by almost $3 \%$. In addition, the anion changes
dramatically the results: For fluorides a large overestimation is obtained for
all PBE$(\beta,\mu,\lambda)$ functionals, while for chlorides (and even more
for bromides, not included here) this behaviour is corrected or some
underestimation can be found.
The change to a tighter Lieb-Oxford bound has a rather strong effect (often
stronger than a reduction of $\mu$) and 
reduces most errors of the IA-VIIA compounds. While
nearly all PBE$(\beta,\mu,\text{LO})$ functionals overestimate the lattice
constants, a tighter Lieb-Oxford bound may lead to small underestimations for
some chlorides. Nevertheless
PBE$(J_r,G_x,\text{EL})$ is the best performing PBE$(\beta,\mu,\lambda)$
functional for this class of compounds.

For the IIA-VIA compounds standard PBE overestimates the lattice
constants. A reduction from $\beta = G_c$ to $\beta = J_s$ at
$\mu = J_r$ [comparing PBE with PBE$(J_s,J_r,\text{\text{LO/EL}})$] improves
PBE significantly, because the change of $\beta$ reduces the effective $\mu$
and therefore PBE$(J_s,J_r,\text{\text{LO}})$ becomes the most accurate
functional for this group of compounds.
The reduction from $\mu = J_r$ to $\mu = G_x$ at $\beta = G_c$ [e.g., from PBE
to PBE$(G_c,G_x,\text{LO})$] also lowers the
lattice constants significantly leading to quite well performing functionals with
$\lambda_{\text{LO}}$ , but with $\lambda_{\text{EL}}$ the correction overshoots and
leads to some underestimation for MgS and CaO. For a fixed $\mu$ the variation
of $\beta$ hardly modifies the results.
Only MgO [which is overestimated by all
PBE$(\beta,\mu,\lambda)$ functionals] benefits from
a tighter Lieb-Oxford bound in all cases.

All semiconducting IIIA-VA compounds have significantly too large lattice
parameters with PBE.
A tighter Lieb-Oxford bound is advantageous, but the effect alone is too small
and a strong reduction of $\mu$ (to $G_x$) is essential. 
For $\mu = G_x$ the parameter for correlation has a fairly large effect (in
particular for the heavier elements) and with $\beta = G_c$ the
PBE$(G_c,G_x,\text{LO})$ functional is very accurate for all semiconductors.

The metallic transition metal compounds (mainly carbides and nitrides, but also
three intermetallic compounds) are fairly well described by standard PBE and
the slight overestimation of lattice constants can be reduced by weak
modifications. The best functionals are obtained either by switching $\lambda$ to the
tighter EL limit [PBE$(G_c,J_r,\text{EL})$], or by a modest reduction of $\beta$
[PBE$(J_s,J_r,\text{LO})$] (probably because
this reduces also the effective $\mu$ for $\mu=J_s$). A stronger reduction of
$\mu$ or a combination of $\mu$ and $\lambda$ reductions leads to too small
lattice parameters.

\section{\label{molecules}Atomization energy of molecules}

\begin{figure}
\includegraphics[width=\columnwidth]{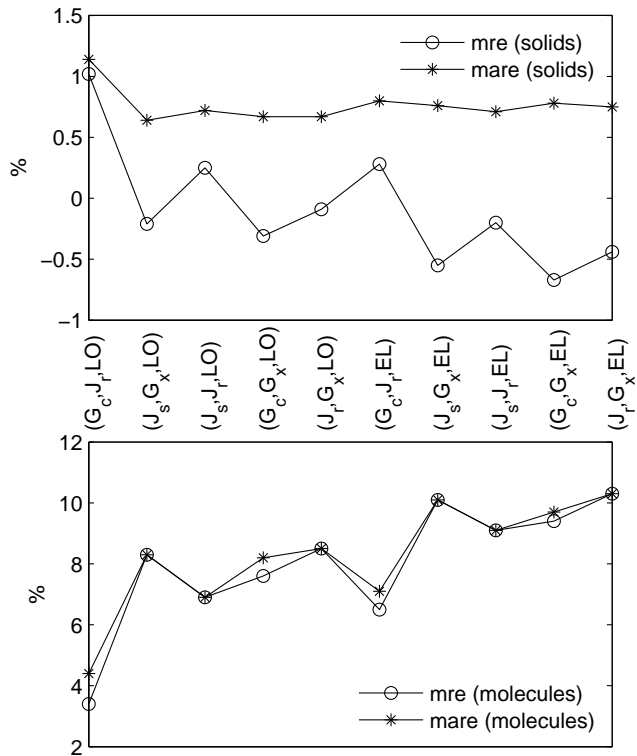}
\caption{\label{fig3} Mean relative error and mean absolute
relative error in the lattice constants (upper panel)
and atomization energies (lower panel) of the PBE$(\beta,\mu,\lambda)$ functionals.}
\end{figure}

In this section we present the performance of the
PBE$(\beta,\mu,\lambda)$ functionals on the atomization energies of
molecules using the representative AE6 test set.\cite{LynchJPCA03} 
Table SII of the supplementary EPAPS material\cite{EPAPS} gives the calculated
values and Table \ref{table2} gives a summary of the corresponding statistical errors.
The me and mae (mre and mare) quantities are very similar since all functionals
(maybe except original PBE) always overestimate the atomization energy.
We can see that
PBE$(G_c,J_r,\text{LO})$ (original PBE) is the best and
PBE$(J_r,G_x,\text{EL})$ is the worst
functional of the PBE$(\beta,\mu,\lambda)$ family for the atomization energy of
molecules.
Switching $\lambda$ to a tighter bound ($\lambda =\lambda_{\text{EL}}$)
has a rather strong degrading
effect, probably because the atomization energies depend a lot on regions in
space with large effective density gradient $s$.
In Fig. \ref{fig3} we compare the mre and mare for
solids (lattice constants) and molecules (atomization energies) versus the
PBE$(\beta,\mu,\lambda)$ functionals. As expected the mre behavior of
solids is opposite to that of the molecules. Functionals leading to larger
lattice constants (larger overestimation) lead 
to smaller atomization energies (smaller overestimation) and no functional of
the current PBE$(\beta,\mu,\lambda)$ family can describe both quantities
in a satisfying way.

\section{Conclusions}
\label{conclusions}

From all of the above we conclude that to obtain precise lattice constants of
solids it is not necessary (and in some cases even detrimental) to switch
from the PBE family of functionals (differing from original PBE only through
the choice of parameters) to functionals that also differ in the form of the
exchange enhancement factor (SOGGA and WC) or that employ different
design principles (AM05) or further ingredients (meta-GGA TPSS). These more 
complex functionals have many merits and interesting features, but apparently
these features are not required to produce very accurate lattice constants.

In fact, even the simplest possible modification of original PBE, the change
of one single parameter (taking $\mu$ from the gradient expansion for
exchange instead of from the jellium response function) already produces a
functional whose lattice constants are, to within the error bars of the
WIEN2k code, as good or better than those of any of the other tested 
functionals: PBE$(G_c,G_x,\text{LO})$. (As pointed out above, Siesta and Gaussian 
calculations sustain this claim.)

A change of two parameters relative to PBE produces PBE$(J_r,G_x,\text{LO})$ and
$\text{PBEsol}=\text{PBE}(J_s,G_x,\text{LO})$, the former having the same mare
but a lower spread,
relative to PBE$(G_c,G_x,\text{LO})$; and the latter a still slightly lower mare
(although the improvement is smaller than our estimated error bar) at the 
expense of a slightly larger spread. Any of these three functionals can be recommended 
as a useful and reliable GGA for lattice constants of solids, requiring only 
minimal changes to existing implementations of PBE and attaining much higher 
accuracy than PBE and LDA, and also than many more complex functionals.

On the other hand, any modification of the original PBE functional which
improves lattice parameters of solids, increases significantly the error in the
atomization energy of molecules. The overestimation of this quantity by PBE of
$\text{mae} = 15.1$ kcal/mol becomes 2$-$3 times larger with the modified functionals
and we must conclude that (at least) GGAs of PBE-form cannot describe well 
lattice parameters of solids and atomization energies of molecules
simultaneously (see also Refs. \onlinecite{sogga} and \onlinecite{PerdewPRL06}).
 
\begin{acknowledgments}

This work was supported by the project P20271-N17 of the Austrian Science Fund
and by the Brazilian funding agencies FAPESP and CNPq.

\end{acknowledgments}

\end{document}